\documentclass[12pt,a4paper]{article}
\usepackage[utf8]{inputenc}
\usepackage[english]{babel}
\usepackage[left=2cm,right=2cm,top=2cm,bottom=2cm]{geometry}

\usepackage{epsf,epsfig,psfrag,graphics,epstopdf}

\usepackage{float}

\usepackage{xcolor}

\usepackage{amsmath}
\usepackage{amsfonts}
\usepackage{amssymb}
\usepackage{graphicx}

\usepackage{cite}

\usepackage{slashed}

\usepackage{color}
\definecolor{nicered}{rgb}{0.7,0.1,0.1}
\definecolor{nicegreen}{rgb}{0.1,0.5,.1}

\usepackage{caption}
\usepackage{subcaption}

\usepackage{hyperref}
\hypersetup{colorlinks,citecolor=nicegreen,linkcolor=nicered}
\hypersetup{colorlinks=true}

\begin{document}




\begin{center}
{\Large \bf Update of the extraction of the CKM matrix elements} \\

\vspace{2mm}

{\small prepared for the proceedings of the 10th international workshop on the CKM Unitarity Triangle} \\

\vspace{4mm}

{\it Luiz Vale Silva, on behalf of the \textsf{\textit{CKM}fitter} Collaboration} \\

\vspace{1mm}

{\small \it University of Sussex, Department of Physics and Astronomy, Falmer, Brighton BN1 9QH, UK} \\

\vspace{1mm}

\texttt{l.h.vale-silva@sussex.ac.uk}
\end{center}

\vspace{0.5cm}

\noindent \textbf{Abstract.}
I discuss the combination and extraction of the Cabibbo-Kobayashi-Maskawa (CKM) matrix elements under the Standard Model (SM) framework from a global fit. The analysis shown here relies on the \textsf{\textit{CKM}fitter} package, consisting in a frequentist approach that employs the \textit{Range} fit (\textit{R}fit) scheme to handle theoretical uncertainties.


\section*{Introduction}


Flavor changing processes provide a powerful test of the SM, and consist, therefore, in an avenue to look for extensions of it.
The breaking of flavor symmetry in the SM is sourced by the Yukawa couplings of the SM Higgs, which give origin to the spectrum of fermion masses and the CKM matrix in the quark sector. The CKM matrix in the SM is a three by three unitary matrix that can be parameterized by three mixing angles and one single source of CP violation.
The CKM matrix turns out to be hierarchical, namely, elements closer to the diagonal are larger. A useful, rephasing invariant, parameterization that shows clearly this feature is the Wolfenstein parameterization consisting of the real parameters $ A $, $ \lambda $, $ \bar\rho $ and $ \bar\eta $

\begin{equation}
	{\color{black} \lambda} = \frac{\vert V_{us} \vert}{(\vert V_{ud} \vert^2 + \vert V_{us} \vert^2)^{1/2}} \, , \quad {\color{black} A \lambda^2} = \frac{\vert V_{cb} \vert}{(\vert V_{ud} \vert^2 + \vert V_{us} \vert^2)^{1/2}} \, , \quad {\color{black} \bar\rho+i \bar\eta} = - \frac{V^{}_{ud} V^{*}_{ub}}{V^{}_{cd} V^{*}_{cb}} \,,
\end{equation}
where no assumption has been made about the size of $ \lambda $, and a small $ \lambda $ corresponds to the hierarchical structure mentioned above.
A useful graphical representation of the CKM matrix results from its unitarity: see, for instance, Figure~\ref{fig:fig1} below, displaying the $ B_d $ triangle.

A rich variety of processes is used to probe the structure of the CKM matrix \cite{Charles:2004jd,Charles:2015gya,Koppenburg:2017mad}, including processes dominated by SM tree-level contributions, processes for which in the SM there is no tree-level contribution, processes that individually rule out a vanishing CP violating phase, or yet processes that are individually compatible with no CP violation. Some observables are presently dominated by experimental uncertainties, while others are dominated by systematic sources of uncertainty,
the latter being due to hadronic effects inherent to quark processes.
We heavily rely on Lattice QCD extractions of hadronic inputs, e.g., bag parameters, decay constants and form factors.

Thanks to the experimental (e.g., $ B $-factories, LHCb, etc.), and theoretical (e.g., Lattice QCD) developments over the last two decades, a much better accuracy has been reached in the extraction of the CKM matrix elements. Here, we briefly discuss the 2018 update (following the conference ICHEP~2018) of the extraction of the CKM matrix elements by the \textsf{\textit{CKM}fitter} Collaboration~\cite{Charles:2004jd}.
Our primary goal is to test the SM, and possibly point out tensions in its description of flavor transitions, likely to occur in presence of physics beyond the SM.

\begin{figure}
	\centering
	\includegraphics[scale=0.5,trim={0 0.1cm 0 0.7cm},clip]{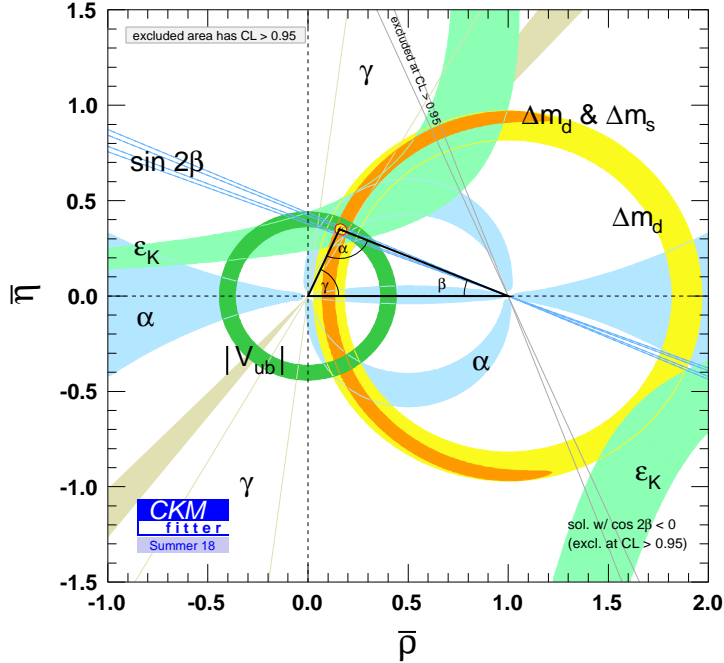}
	\caption{Combination of constraints used in the extraction of the Wolfenstein parameters in the $ ( \bar{\rho} \, , \bar{\eta} ) $ plane. The apex of the triangle at $ \bar{\rho} \simeq 0.16 $ and $ \bar{\eta} \simeq 0.35 $ corresponds to the best-fit point. Also shown are the $ 68~\% $ (hashed red) and $ 95~\% $ (yellow with red contour) Confidence Level regions of the true value of $ \bar{\rho} $ and $ \bar{\eta} $.}\label{fig:fig1}
\end{figure}

\section*{Statistical approach}


The \textsf{\textit{CKM}fitter} Collaboration employs a frequentist framework based on a $ \chi^2 $ analysis. The approach used to incorporate and manipulate systematic uncertainties is called \textit{Range} fit (\textit{R}fit) scheme. In practice, it means that one can vary freely the true value of the fixed (and thus not stochastic) and unknown theoretical correction $ \delta $ inside the quoted uncertainty $ \pm \Delta $, i.e., $ \delta \in [- \Delta, \Delta] $, without any penalty from the $ \chi^2 $. This setup implies a plateau for the preferred value of the parameter carrying the theoretical uncertainty $ \pm \Delta $. Subsequently, Confidence Level (C.L.) intervals are determined from the variation of the $ \chi^2 $ around the best-fit point.


As previously stated,
important systematic, or theoretical, uncertainties are usually present in the quark flavor sector.
While here we adopted the \textit{R}fit scheme introduced above, other approaches are discussed and compared in Ref.~\cite{Charles:2016qtt}.
Though their treatment is somewhat ill-defined, a useful and meaningful modeling of systematic uncertainties must satisfy a certain number of criteria, such as good coverage properties (at least for the confidence level significances we are interested in), propagation of uncertainties (i.e., a clear separation of statistical and theoretical uncertainties), and the more technical requirement of being simple enough to allow for tractable calculations (i.e., the determination of the best-fit point and confidence intervals) when dealing with a large number of constraints and parameters to be extracted. See, e.g., Ref.~\cite{Charles:2016qtt} for a more detailed discussion.




The Lattice QCD inputs used here, assessing hadronic quantities such as bag parameters, decay constants, form factors, etc., come from published references only (including proceedings), with uncertainty budgets provided, and consist of unquenched results with $ 2 $, $ 2+1 $, or $ 2+1+1 $ dynamical fermions.
We follow the \textsf{FLAG} reviews \cite{Aoki:2016frl} in order to keep track of relevant Lattice QCD extractions.
If for a quantity we have many sources of systematic uncertainty, they are treated at the same footing and, in the absence of correlations, summed linearly (see Ref.~\cite{Charles:2016qtt} for a discussion of the definition, and inclusion, of correlations among theoretical uncertainties in the same spirit of the \textit{R}fit model for uncertainties).
Different extractions of the same quantity are combined following a conservative procedure called \textit{educated Rfit} in which the averaged theoretical uncertainty is not smaller than the smallest of the individual theoretical uncertainties. More details are provided at Refs.~\cite{Charles:2004jd,Charles:2015gya,Charles:2016qtt}.







\section*{Results and discussion}

\begin{figure}
	\centering
	\includegraphics[scale=0.395,trim={3cm 0.5cm 0.5cm 0.5cm},clip]{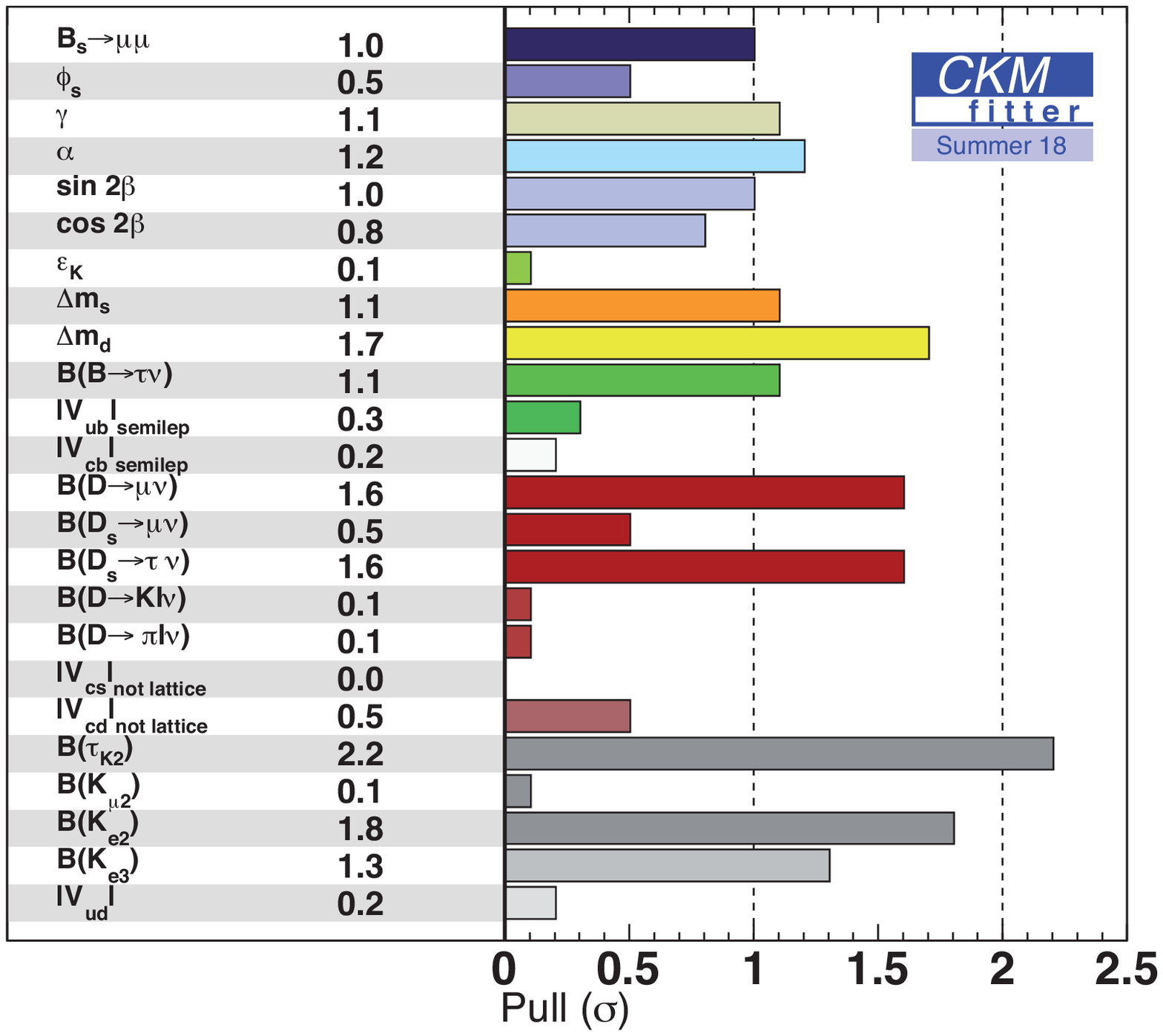} \hspace{9mm} \includegraphics[scale=0.39,trim={0.5cm 0.5cm 0.5cm 0.5cm},clip]{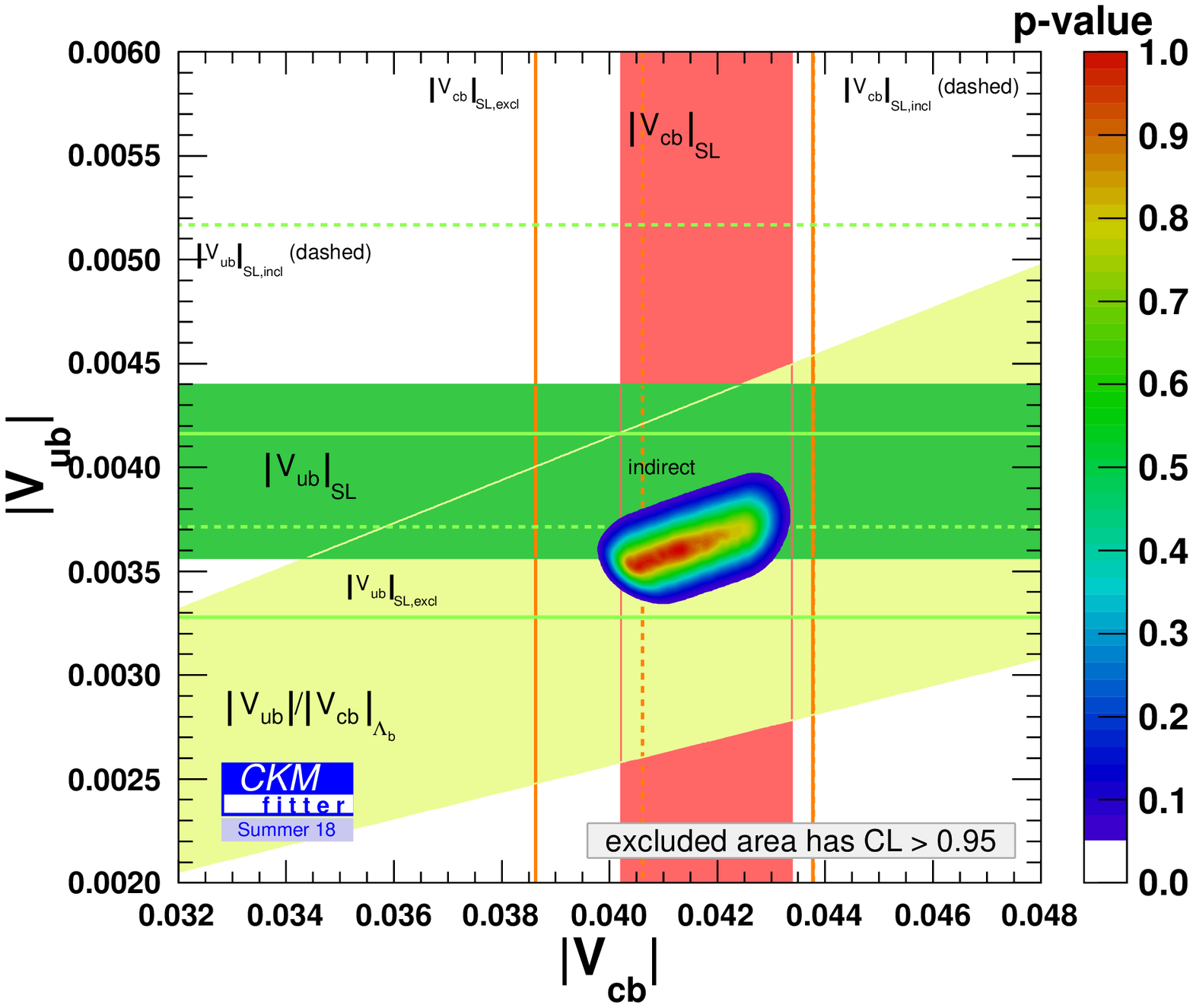}
	\caption{{\color{nicegreen} \it Left panel.} Pulls for individual observables in units of $ \sigma $. {\color{nicegreen} \it Right panel.} Status of the $ ( | V_{cb} | \, , | V_{ub} | ) $ plane. The horizontal and vertical coloured bands represent our averages of the determinations from semileptonic $ B $ decays. The white bands with solid (dashed) borders correspond to the determinations from exclusive (inclusive) semileptonic $ B $ decays. The diagonal coloured band corresponds to the determination of $ | V_{ub} |/| V_{cb} | $ from $ \Lambda_b $ decays. The rainbow oval region indicates the indirect determination of $ | V_{ub} | $ and $ | V_{cb} | $ from the global fit, without any information from semileptonic or leptonic decays of $ b $-hadrons.}\label{fig:fig2}
\end{figure}


The full set of experimental and theoretical inputs is given at Ref.~\cite{website}.
Compared to our Summer 2016 edition (following the conference ICHEP~2016), the results of the global fit under the SM hypothesis remain excellent: the p-value is $ 51\% $, which corresponds to $ 0.7\sigma $, if all uncertainties are treated as Gaussian.
The consistent overall picture allows for a meaningful extraction of the CKM matrix elements, the extracted Wolfenstein parameters being ($ 68\% $ C.L. intervals)



\begin{eqnarray}\label{eq:fullGlobalFitExtraction}
A = 0.8403^{\,+0.0056}_{\,-0.0201} \, (2\% \, {\rm unc.})\,, &\qquad&
\lambda = 0.224747^{\,+0.000254}_{\,-0.000059} \, (0.07\% \, {\rm unc.})\,,\\
\bar\rho = 0.1577^{\,+0.0096}_{\,-0.0074} \, (5\% \, {\rm unc.})\,, &\qquad&
\bar\eta = 0.3493^{\,+0.0095}_{\,-0.0071} \, (2\% \, {\rm unc.})\,.\nonumber
\end{eqnarray}
In the global fit analysis, the small parameter $ \lambda $ is accurately determined from  $ | V_{ud} | $ (superallowed nuclear transitions) and $ | V_{us} | $ (semileptonic kaon decays), while the parameter $ A $ is accurately determined from a combination of $ | V_{cb} | $, $ | V_{ub} | $, and other information from the global fit. The extraction of $ \bar{\rho} $ and $ \bar{\eta} $ is illustrated in Figure~\ref{fig:fig1}, where it is clear the dominant role played by $ \sin (2 \beta) $ and $ \Delta m_{d} / \Delta m_{s} $, but also $ | V_{ub} | $, in the determination of the CP violating phase. For completeness,
the Wolfenstein parameters extracted from a global fit including only observables dominated by SM tree-level contributions, for which the p-value computed with Gaussian uncertainties is $ 43\%~(0.8\sigma) $, are ($ 68\% $ C.L. intervals)


\begin{eqnarray}
A = 0.8396^{\,+0.0080}_{\,-0.0298} \, (2\% \, {\rm unc.})\,, &\qquad&
\lambda = 0.224756^{\,+0.000163}_{\,-0.000065} \, (0.05\% \, {\rm unc.})\,,\nonumber\\
\bar\rho = 0.123 \pm 0.023 \, (19\% \, {\rm unc.})\,, &\qquad&
\bar\eta = 0.375^{\,+0.022}_{\,-0.017} \, (5\% \, {\rm unc.})\,,\nonumber
\end{eqnarray}
in good agreement with the extraction from the full global fit of Eq.~\eqref{eq:fullGlobalFitExtraction}.




\begin{figure}
	\centering
	\includegraphics[scale=0.27]{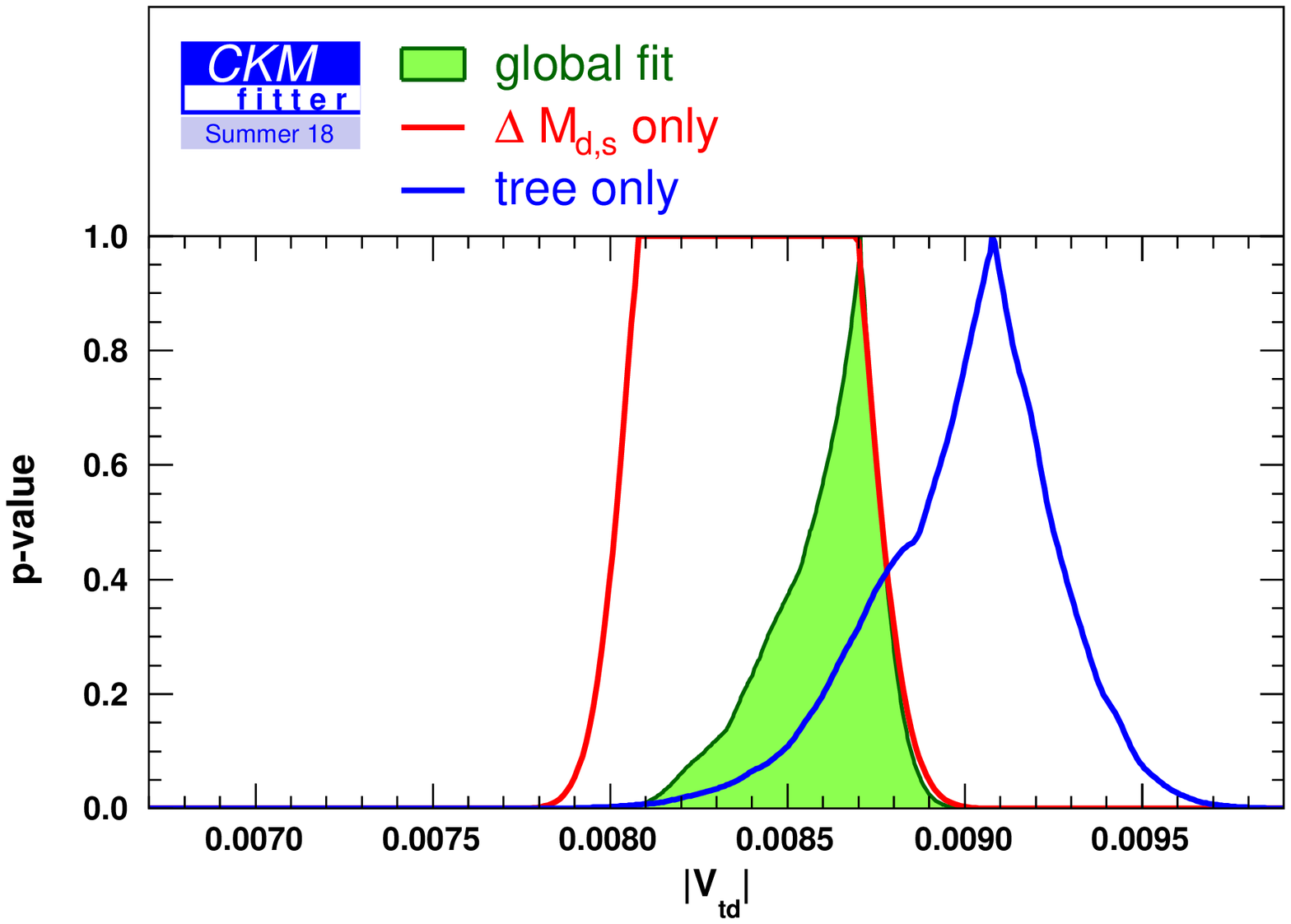}
	\hspace{1mm} \includegraphics[scale=0.27]{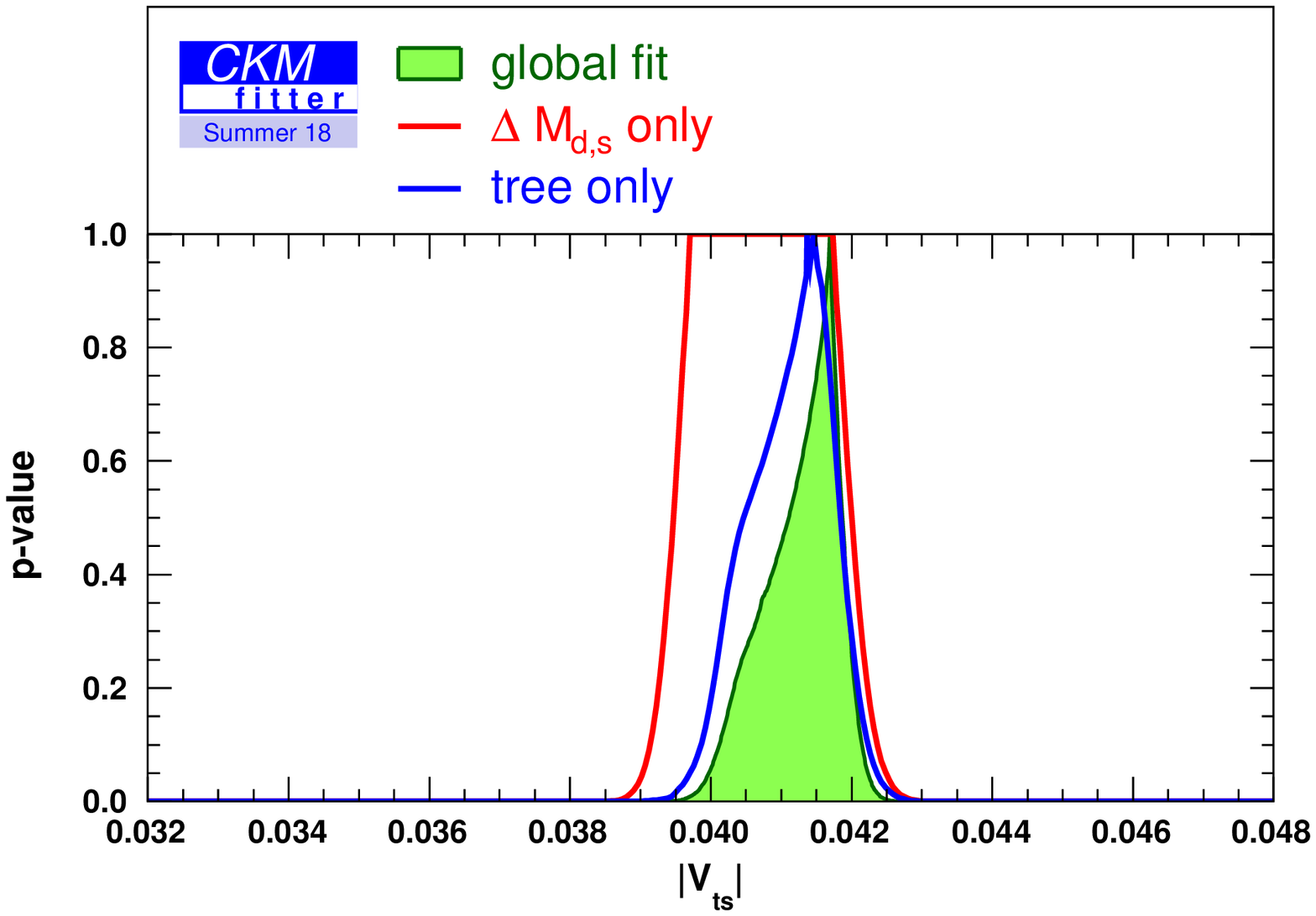}	
	\hspace{1mm} \includegraphics[scale=0.27]{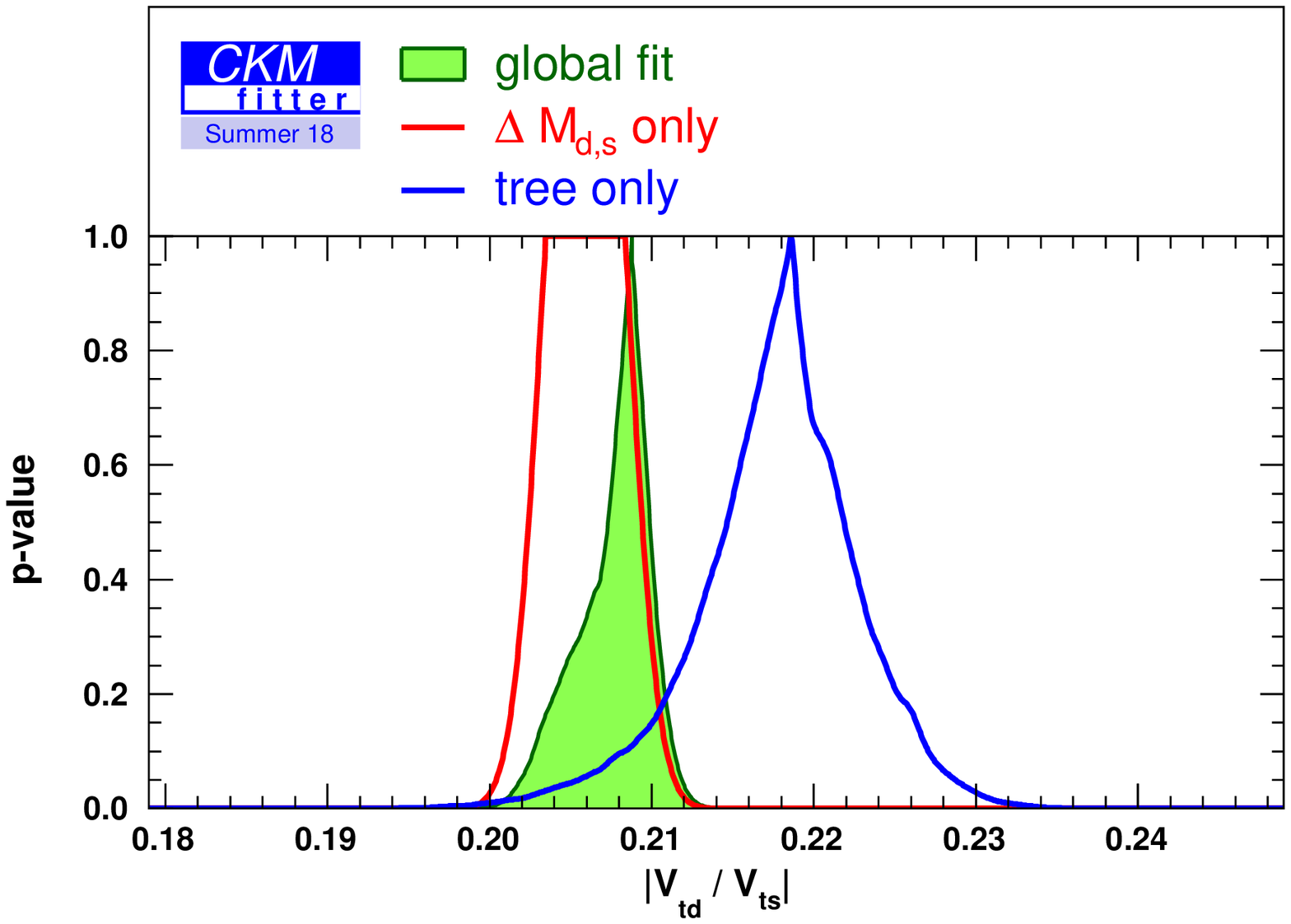}
	\caption{Constraint on $ | V_{td} | $ ({\color{nicegreen} \it left}), $ | V_{ts} | $ ({\color{nicegreen} \it center}), and $ | V_{td} |/| V_{ts} | $ ({\color{nicegreen} \it right}) from: the global fit (green), using only information on $ B_d $ and $ B_s $ mixings (red), and using only tree-level quantities (blue).}\label{fig:fig3}
\end{figure}



Each observable considered in the fit provides a test of the validity of the SM, as measured by the pulls defined by

\begin{equation}
	Pull_{\mathcal{O}_{}} = \sqrt{\chi^2_{min} - \chi^2_{min, !\mathcal{O}_{}}} \,,
\end{equation}
where $ !\mathcal{O}_{} $ means that the corresponding $ \chi^2 $ is built without the input for $ \mathcal{O}_{} $.
The values of the pulls given in the left panel of Figure~\ref{fig:fig2} imply that the SM predictions are in good agreement with the measurements. Since the observables have correlated fits (e.g., the $ ( | V_{cb} | \, , | V_{ub} | ) $ plane shown in the right panel of Figure~\ref{fig:fig2}) and since theoretical uncertainties are not treated as Gaussian, the distribution of pulls is not normal, and the number of observables which have a pull larger than $ n \times \sigma $, for a certain $ n $, is not always a meaningful information.
Note that the presence of a plateau in the \textit{R}fit model for theoretical uncertainties may lead to a vanishing pull for some quantities even in cases where the predicted and the observed values are not identical.










We now discuss semileptonic extractions of $ | V_{cb} | $, for which there has been a long-standing tension between inclusive and exclusive extractions.
Recent studies show discrepancies in the exclusive extractions of $ | V_{cb} | $ following the use of CLN \cite{Caprini:1997mu} or BGL \cite{Boyd:1997kz} parameterizations of the form factors, which may be due to residual uncertainties in the former approach. Here, we employ only the BGL parameterization. We have considered the following inputs for exclusive extractions:
$ |V_{cb}|_{B \to D} = (40.5 \pm 0.8 ({\rm exp.}) \pm 0.6 ({\rm LQCD}) \pm 0.2 ({\rm QED}))\times 10^{-3} $, from combined Belle and BaBar data \cite{Lattice:2015rga,Bigi:2016mdz} (see also references therein); and
$ |V_{cb}|_{B \to D^*} = (42.4 \pm 0.7 ({\rm exp.}) \pm 1.1 ({\rm LQCD}) \pm 0.2 ({\rm QED})) \times 10^{-3} $, from combined Belle tagged and untagged data \cite{Bailey:2014tva,Grinstein:2017nlq,Abdesselam:2017kjf,Abdesselam:2018nnh}; they lead to the following average


\begin{equation}
	|V_{cb}|_{excl.}=( 41.2\pm0.6({\rm exp.}) \pm0.9({\rm LQCD}) \pm0.2({\rm QED})  ) \times 10^{-3} \,.
\end{equation}
Following Ref.~\cite{Bailey:2014tva}, an additional QED uncertainty has been systematically included. For the inclusive extraction, we have considered $ |V_{cb}|_{incl.}=( 42.2\pm0.4 \pm0.6  ) \times 10^{-3} $ \cite{Gambino:2013rza,Amhis:2016xyh,website}. Note that the inputs for exclusive and inclusive values are in good agreement. Their combination leads to the following average:


\begin{equation}
	|V_{cb}|=( 41.8 \pm 0.4 \pm 0.6  ) \times 10^{-3} \,,
\end{equation}
which is about $ 1\sigma $ higher than our 2016 value (namely, $|V_{cb}|=( 41.0 \pm 0.3 \pm 0.7 ) \times 10^{-3}$).
The status of $ | V_{cb} | $, together with $ | V_{ub} | $, is shown in the right panel of Figure~\ref{fig:fig2}. As already mentioned, there is no tension between the values of $ |V_{cb}|_{excl.} $ and $ |V_{cb}|_{incl.} $ given above; the fit still favors the exclusive extraction of $ | V_{ub} | $, which remains somewhat in tension with the inclusive one.




Finally, we discuss different extractions of $ | V_{td} | $ and $ | V_{ts} | $. We consider extractions from: tree-level only processes, by relying on the unitarity of the CKM matrix, and from $ B_{(s)} $ mixings, which require decay constants and bag parameters as theoretical inputs. The latter are taken from Refs.~\cite{Carrasco:2013zta,Aoki:2014nga,Bazavov:2016nty} and averaged following the \textit{R}fit scheme.
As seen from Figure~\ref{fig:fig3}, the extraction from tree-level only processes favors larger values of $ | V_{td} | $ and $ | V_{td} |/| V_{ts} | $ compared to the one from $ B_{(s)} $ mixings only. Though some discrepancy between both extractions is present, it does not reach important levels of significance.

\vspace{0.2cm}

In conclusion, the SM gives a consistent picture of flavor transitions and CP violation in the quark sector. At the moment, the global fit does not point to large deviations from the SM (see Ref.~\cite{future} for future prospects concerning searches for NP in $ |\Delta F| = 2 $ processes). More information relative to the 2018 edition discussed here is found at our website, Ref.~\cite{website}.
The reader is invited to use the web interface Ref.~\cite{ckmlive} to prepare her/his own global fit analyses.




\bibliography{mybib}{}
\bibliographystyle{unsrt}

\end{document}